\documentclass[aps,prd,floats,floatfix,nofootinbib,superscriptaddress]{revtex4-1}
\usepackage{graphicx}
\usepackage{amssymb}
\usepackage{hyperref}

\usepackage{amsmath}
\usepackage{bbold}
\usepackage{color}

\begin{document}

\title{A benchmark for LHC searches for low-mass custodial fiveplet scalars in the Georgi-Machacek model}

\author{Ameen Ismail}
\email{ai279@cornell.edu}
\affiliation{Department of Physics, LEPP, Cornell University, Ithaca, New York 14853, USA}

\author{Ben Keeshan}
\email{BenKeeshan@cmail.carleton.ca}
\affiliation{Ottawa-Carleton Institute for Physics, Carleton University, 1125 Colonel By Drive, Ottawa, Ontario K1S 5B6, Canada}

\author{Heather E.~Logan}
\email{logan@physics.carleton.ca} 
\affiliation{Ottawa-Carleton Institute for Physics, Carleton University, 1125 Colonel By Drive, Ottawa, Ontario K1S 5B6, Canada}

\author{Yongcheng Wu}
\email{ycwu@physics.carleton.ca}
\affiliation{Ottawa-Carleton Institute for Physics, Carleton University, 1125 Colonel By Drive, Ottawa, Ontario K1S 5B6, Canada}

\date{March 4, 2020}                                           

\begin{abstract}
The Georgi-Machacek (GM) model is used to motivate and interpret LHC searches for doubly charged scalars decaying to vector bosons pairs. The doubly charged scalars are part of a degenerate fermiophobic custodial fiveplet with states $H_5^{\pm\pm}$, $H_5^\pm$, and $H_5^0$ and common mass $m_5$. The GM model has been extensively studied at the LHC for $m_5 > 200$~GeV but there is a largely unprobed region of parameter space from $120$~GeV $< m_5 < 200$~GeV where light doubly-charged scalars could exist. This region has been neglected by experimental searches due in part to the lack of a benchmark for $m_5 < 200$~GeV. In this paper we propose a new ``low-$m_5$'' benchmark for the GM model, defined for $m_5 \in (50, 550)$~GeV, and characterize its properties.  We apply all existing experimental constraints and summarize the phenomenology of the surviving parameter space.  We show that the benchmark populates almost the entirety of the relevant allowed parameter plane for $m_5$ below 200~GeV and satisfies the constraints from the 125~GeV Higgs boson signal strengths. We compute the 125~GeV Higgs boson's couplings to fermion and vector boson pairs and show that they are always enhanced in the benchmark relative to those in the Standard Model. We also compute the most relevant production cross sections for $H_5$ at the LHC, including Drell-Yan production of $H_5$ pairs.  The process $pp \to H \to H_5 H_5$ contributes in a small region of parameter space, but is small compared to the Drell-Yan production cross section. Finally we compute the branching ratios of $H_5^0 \to \gamma\gamma$ and $H_5^{\pm} \to W^{\pm} \gamma$.  The width-to-mass ratio of each of the $H_5$ states is below 1\% over the entire benchmark, so that the narrow-width approximation is well justified.
\end{abstract}

\maketitle 

\section{Introduction}

The Georgi-Machacek (GM) model~\cite{Georgi:1985nv,Chanowitz:1985ug} provides a prototype for models in which part of electroweak symmetry breaking is due to the vacuum expectation value (vev) of scalars in SU(2)$_L$ representations larger than the doublet (``exotic'' scalars).  These models have as a common feature the possibility that the Standard Model (SM)-like Higgs boson $h$ can couple to $W$ and $Z$ boson pairs with a strength greater than that in the SM~\cite{Georgi:1985nv,Falkowski:2012vh,Hisano:2013sn,Kanemura:2013mc,Logan:2015xpa}, due to enhancement factors arising from the SU(2)$_L \times$U(1)$_Y$ generators in the gauge-kinetic terms of the scalar Lagrangian.  They include generalizations of the GM model to higher isospin~\cite{Logan:2015xpa} and the scalar septet model~\cite{Hisano:2013sn,Kanemura:2013mc}.  These models also contain as a common feature singly- and doubly-charged Higgs bosons that couple at tree level to vector boson pairs and  play a crucial role in the unitarization of longitudinal vector boson scattering amplitudes in these models~\cite{Falkowski:2012vh}.  In the GM model, these are the so-called custodial-fiveplet $H_5^{\pm\pm}$, $H_5^{\pm}$, and $H_5^0$ states, which have a common mass $m_5$ and are fermiophobic.  Their coupling to vector boson pairs is parameterized by $s_H \equiv \sin\theta_H = 2 \sqrt{2} v_\chi/v$, where $v_\chi$ is the vev of the isospin-triplet scalars in the GM model and $v$ is the Higgs vev in the SM. In particular, $s_H^2$ is equal to the fraction of the $W$ and $Z$ boson squared-masses that is generated by the triplet vevs. Assuming that the branching ratio of the $H_5$ to vector bosons is close to 1, the proportionality of the vector boson fusion production cross section of the $H_5$ states to $s_H^2$ allows LHC searches to directly constrain $s_H$ as a function of $m_5$ ~\cite{Aad:2015nfa,Sirunyan:2017sbn,Sirunyan:2017ret,Aaboud:2018ohp,CMS:2018ysc,Aaboud:2017gsl,Aaboud:2018qcu}.      

The current LHC searches for the $H_5$ states in the GM model focus on $H_5$ masses of 200~GeV and above, for which the decays into vector boson pairs are on shell. Below this mass, there exists a largely unprobed region in the parameter space with doubly-charged Higgs bosons as light as 120~GeV still allowed. However, this open region of parameter space could easily be constrained by extending existing experimental searches to lower masses. For example, the ATLAS search for Drell-Yan production of $H^{++}H^{--}$ with decays to like-sign $W$ bosons~\cite{Aaboud:2018qcu}, if extended to masses below 200~GeV and using the full Run 2 data-set, would probe this entire neglected low-mass region independent of $s_H$ and could potentially entirely exclude $H_5$ masses below 200--300~GeV. (The only possible loophole to such an exclusion would be mass spectra engineered to make the Higgs-to-Higgs decays $H_5^{++} \to H_3^+ W^+$ and/or $H_5^{++} \to H_3^+ H_3^+$ dominate over the $H_5^{++} \to W^+W^+$ decay mode.  It is not yet known whether such a scenario can be achieved while satisfying all theoretical and experimental constraints.) 
To simplify the interpretation of the above searches, the LHC Higgs Cross Section Working Group developed the H5plane benchmark scenario for the GM model~\cite{deFlorian:2016spz,Logan:2017jpr}, defined for $m_5 \in [200, 2000]$~GeV. The H5plane benchmark takes as its two free parameters $m_5$ and $s_H$ and the other parameters are fixed to ensure $BR(H_5 \to VV) = 1$ to a good approximation. However, for $m_5 < 200$~GeV, the H5plane benchmark is unable to cover a significant fraction of the region allowed in a general parameter scan.

In this paper we propose and study a new benchmark plane in the GM model that is valid in the low-$m_5$ region, defined in our case to be $m_5 \in (50, 550)$~GeV. The purpose of this benchmark is to facilitate experimental searches that extend to $m_5$ values below 200~GeV. This benchmark is designed to yield $BR(H_5 \to VV) = 1$ while avoiding constraints from the measurements of the 125~GeV Higgs boson signal strengths; in particular, the parameters are chosen to avoid large modifications to $h \to \gamma\gamma$ from loops of light $H_5^{\pm}$ and $H_5^{\pm\pm}$.  We show that the low-$m_5$ benchmark successfully populates a large portion of the theoretically allowed region and almost the entire experimentally allowed region. We also show that the SM-like Higgs couplings to fermion and vector boson pairs, $\kappa_f^h$ and $\kappa_V^h$, are always enhanced in the benchmark with $\kappa_f^h$ as large as 1.19 and $\kappa_V^h$ as large as 1.08.  We compute the cross sections for Drell-Yan production of pairs of $H_5$ states as well as their single production cross sections via vector boson fusion (VBF).  We also point out a small region of parameter space in which $H_5$ pairs can be produced via on-shell decays of the heavier custodial-singlet scalar $H$, though the cross sections for this production mode are more than an order of magnitude smaller than those from Drell-Yan. We present the branching ratios of the loop suppressed decays of the $H_5^0$ and $H_5^+$ which become significant at low values of $m_5$ and highlight that the dominant exclusion in the $m_5 <$ 200~GeV region comes from $H_5^0 \to \gamma \gamma$ which limits this branching ratio to be less than approximately 5\%. Finally, we note that the $H_5$ width to mass ratios are below 1\%, which is sufficiently small to justify the narrow-width approximation. 

The structure of the paper is as follows. In Section \ref{sec:model} we present the specifics of the GM model. In Section \ref{sec:bench} we introduce the low-$m_5$ benchmark. In Section \ref{sec:pop} we compare the low-$m_5$ benchmark to a general parameter scan and discuss the relevant experimental constraints. In Section \ref{sec:pheno} we discuss the phenomenological characteristics of the low-$m_5$ benchmark plane, focusing on the couplings of the 125~GeV Higgs boson, the $H_5$ production cross-sections, and the dominant decays of the $H_5$ states. In Section \ref{sec:conc} we conclude.

\section{Georgi-Machacek model}
\label{sec:model}

The scalar sector of the Georgi-Machacek model~\cite{Georgi:1985nv,Chanowitz:1985ug} consists of the usual complex doublet $(\phi^+,\phi^0)$ with hypercharge\footnote{We use $Q = T^3 + Y/2$.} $Y = 1$, a real 
triplet $(\xi^+,\xi^0,\xi^-)$ with $Y = 0$, and  a complex triplet $(\chi^{++},\chi^+,\chi^0)$ with $Y=2$.  The doublet is responsible for the fermion masses as in the SM.
In order to avoid stringent constraints on the electroweak $\rho$ parameter, the model enforces a global SU(2)$_L \times$SU(2)$_R$ symmetry in the scalar potential which breaks to the diagonal subgroup, identified with the custodial SU(2), upon electroweak symmetry breaking. 
In order to make the global SU(2)$_L \times$SU(2)$_R$ symmetry explicit, we write the doublet in the form of a bi-doublet $\Phi$ and combine the triplets to form a bi-triplet $X$:
\begin{eqnarray}
\Phi &=& \left( \begin{array}{cc}
\phi^{0*} &\phi^+  \\
-\phi^{+*} & \phi^0  \end{array} \right), \\
X &=&
\left(
\begin{array}{ccc}
\chi^{0*} & \xi^+ & \chi^{++} \\
-\chi^{+*} & \xi^{0} & \chi^+ \\
\chi^{++*} & -\xi^{+*} & \chi^0  
\end{array}
\right).
\label{eq:PX}
\end{eqnarray}
The vevs are defined by $\langle \Phi  \rangle = \frac{ v_{\phi}}{\sqrt{2}} I_{2\times2}$  and $\langle X \rangle = v_{\chi} I_{3 \times 3}$, where $I_{n \times n}$ is the unit matrix and the $W$ and $Z$ boson masses constrain
\begin{equation}
v_{\phi}^2 + 8 v_{\chi}^2 \equiv v^2 = \frac{1}{\sqrt{2} G_F} \approx (246~{\rm GeV})^2,
\label{eq:vevrelation}
\end{equation} 
where $G_F$ is the Fermi constant.
Note that the two triplet fields $\chi^0$ and $\xi^0$ must obtain the same vev in order to preserve the custodial SU(2) symmetry.
Furthermore we will decompose the neutral fields into real and imaginary parts according to
\begin{equation}
\phi^0 \to \frac{v_{\phi}}{\sqrt{2}} + \frac{\phi^{0,r} + i \phi^{0,i}}{\sqrt{2}},
\qquad
\chi^0 \to v_{\chi} + \frac{\chi^{0,r} + i \chi^{0,i}}{\sqrt{2}}, 
\qquad
\xi^0 \to v_{\chi} + \xi^{0,r},
\end{equation}
where we note that $\xi^0$ is already a real field.

The most general gauge-invariant scalar potential involving these fields that preserves custodial SU(2) is given by~\cite{Hartling:2014zca}
\begin{eqnarray}
V(\Phi,X) &= & \frac{\mu_2^2}{2}  \text{Tr}(\Phi^\dagger \Phi) 
+  \frac{\mu_3^2}{2}  \text{Tr}(X^\dagger X)  
+ \lambda_1 [\text{Tr}(\Phi^\dagger \Phi)]^2  
+ \lambda_2 \text{Tr}(\Phi^\dagger \Phi) \text{Tr}(X^\dagger X)   \nonumber \\
& & + \lambda_3 \text{Tr}(X^\dagger X X^\dagger X)  
+ \lambda_4 [\text{Tr}(X^\dagger X)]^2 
- \lambda_5 \text{Tr}( \Phi^\dagger \tau^a \Phi \tau^b) \text{Tr}( X^\dagger t^a X t^b) 
\nonumber \\
& & - M_1 \text{Tr}(\Phi^\dagger \tau^a \Phi \tau^b)(U X U^\dagger)_{ab}  
-  M_2 \text{Tr}(X^\dagger t^a X t^b)(U X U^\dagger)_{ab}.
\label{eq:potential}
\end{eqnarray} 
Here the SU(2) generators for the doublet representation are $\tau^a = \sigma^a/2$ with $\sigma^a$ being the Pauli matrices,
the generators for the triplet representation are
\begin{equation}
t^1= \frac{1}{\sqrt{2}} \left( \begin{array}{ccc}
0 & 1  & 0  \\
1 & 0  & 1  \\
0 & 1  & 0 \end{array} \right), \quad  
t^2= \frac{1}{\sqrt{2}} \left( \begin{array}{ccc}
0 & -i  & 0  \\
i & 0  & -i  \\
0 & i  & 0 \end{array} \right), \quad 
t^3= \left( \begin{array}{ccc}
1 & 0  & 0  \\
0 & 0  & 0  \\
0 & 0 & -1 \end{array} \right),
\end{equation}
and the matrix $U$, which rotates $X$ into the Cartesian basis, is given by~\cite{Aoki:2007ah}
\begin{equation}
U = \left( \begin{array}{ccc}
- \frac{1}{\sqrt{2}} & 0 &  \frac{1}{\sqrt{2}} \\
- \frac{i}{\sqrt{2}} & 0  &   - \frac{i}{\sqrt{2}} \\
0 & 1 & 0 \end{array} \right).
\label{eq:U}
\end{equation}
We note that all the operators in Eq.~(\ref{eq:potential}) are manifestly Hermitian, so that the parameters in the scalar potential must all be real.  Explicit CP violation is thus not possible in the Georgi-Machacek model.  

In terms of the vevs, the scalar potential is given by
\begin{equation}
V(v_\phi,v_\chi) = \frac{\mu_2^2}{2} v_\phi^2 + 3 \frac{\mu_3^2}{2} v_\chi^2
+ \lambda_1 v_\phi^4 
+ \frac{3}{2} \left( 2 \lambda_2 - \lambda_5 \right) v_\phi^2 v_\chi^2
+ 3 \left( \lambda_3 + 3 \lambda_4 \right) v_\chi^4
- \frac{3}{4} M_1 v_\phi^2 v_\chi - 6 M_2 v_\chi^3.
\end{equation}
Minimizing this potential yields the following constraints:
\begin{eqnarray}
0 = \frac{\partial V}{\partial v_{\phi}} &=& 
v_{\phi} \left[ \mu_2^2 + 4 \lambda_1 v_{\phi}^2 
+ 3 \left( 2 \lambda_2 - \lambda_5 \right) v_{\chi}^2 - \frac{3}{2} M_1 v_{\chi} \right], 
\label{eq:phimincond} \\
0 = \frac{\partial V}{\partial v_{\chi}} &=& 
3 \mu_3^2 v_{\chi} + 3 \left( 2 \lambda_2 - \lambda_5 \right) v_{\phi}^2 v_{\chi}
+ 12 \left( \lambda_3 + 3 \lambda_4 \right) v_{\chi}^3
- \frac{3}{4} M_1 v_{\phi}^2 - 18 M_2 v_{\chi}^2.
\label{eq:chimincond}
\end{eqnarray}
Inserting $v_{\phi}^2 = v^2 - 8 v_{\chi}^2$ [Eq.~(\ref{eq:vevrelation})] into Eq.~(\ref{eq:chimincond}) yields a cubic equation for $v_{\chi}$ in terms of $v$, $\mu_3^2$, $\lambda_2$, $\lambda_3$, $\lambda_4$, $\lambda_5$, $M_1$, and $M_2$.  With $v_{\chi}$ (and hence $v_{\phi}$) in hand, Eq.~(\ref{eq:phimincond}) can be used to eliminate $\mu_2^2$ in terms of the parameters in the previous sentence together with $\lambda_1$.  We illustrate below how $\lambda_1$ can also be eliminated in favor of one of the custodial singlet Higgs masses $m_h$ or $m_H$ [see Eq.~(\ref{eq:lambda1})].

The physical field content is as follows.
When expanded around the minimum, the scalar potential gives rise to ten real physical fields together with three Goldstone bosons.  The Goldstone bosons are given by
\begin{eqnarray}
G^+ &=& c_H \phi^+ + s_H \frac{\left(\chi^++\xi^+\right)}{\sqrt{2}}, \nonumber\\
G^0  &=& c_H \phi^{0,i} + s_H \chi^{0,i},
\end{eqnarray}
where
\begin{equation}
c_H \equiv \cos\theta_H = \frac{v_{\phi}}{v}, \qquad
s_H \equiv \sin\theta_H = \frac{2\sqrt{2}\,v_\chi}{v}.
\end{equation}
The physical fields can be organized by their transformation properties under the custodial SU(2) symmetry into a fiveplet, a triplet, and two singlets.  The fiveplet and triplet states are given by
\begin{eqnarray}
H_5^{++}  &=&  \chi^{++}, \nonumber\\
H_5^+ &=& \frac{\left(\chi^+ - \xi^+\right)}{\sqrt{2}}, \nonumber\\
H_5^0 &=& \sqrt{\frac{2}{3}} \xi^{0,r} - \sqrt{\frac{1}{3}} \chi^{0,r}, \nonumber\\
H_3^+ &=& - s_H \phi^+ + c_H \frac{\left(\chi^++\xi^+\right)}{\sqrt{2}}, \nonumber\\
H_3^0 &=& - s_H \phi^{0,i} + c_H \chi^{0,i}.
\end{eqnarray}
Within each custodial multiplet, the masses are degenerate at tree level.  Using Eqs.~(\ref{eq:phimincond}--\ref{eq:chimincond}) to eliminate $\mu_2^2$ and $\mu_3^2$, the fiveplet and triplet masses can be written as
\begin{eqnarray}
m_5^2 &=& \frac{M_1}{4 v_{\chi}} v_\phi^2 + 12 M_2 v_{\chi} 
+ \frac{3}{2} \lambda_5 v_{\phi}^2 + 8 \lambda_3 v_{\chi}^2, \nonumber \\
m_3^2 &=&  \frac{M_1}{4 v_{\chi}} (v_\phi^2 + 8 v_{\chi}^2) 
+ \frac{\lambda_5}{2} (v_{\phi}^2 + 8 v_{\chi}^2) 
= \left(  \frac{M_1}{4 v_{\chi}} + \frac{\lambda_5}{2} \right) v^2.
\label{eq:m5m3}
\end{eqnarray}
Note that the ratio $M_1/v_{\chi}$ is finite in the limit $v_{\chi} \to 0$, as can be seen from Eq.~(\ref{eq:chimincond}) which yields
\begin{equation}
\frac{M_1}{v_{\chi}} = \frac{4}{v_{\phi}^2} 
\left[ \mu_3^2 + (2 \lambda_2 - \lambda_5) v_{\phi}^2 
+ 4(\lambda_3 + 3 \lambda_4) v_{\chi}^2 - 6 M_2 v_{\chi} \right].
\end{equation}

The two custodial SU(2) singlets are given in the gauge basis by
\begin{equation}
\phi^{0,r}, \qquad
H_1^{0 \prime} \equiv \sqrt{\frac{1}{3}} \xi^{0,r} + \sqrt{\frac{2}{3}} \chi^{0,r}.
\end{equation}
These states mix by an angle $\alpha$ to form the two custodial-singlet mass eigenstates $h$ and $H$, defined such that $m_h < m_H$:
\begin{equation}
h = c_{\alpha} \phi^{0,r} - s_{\alpha} H_1^{0\prime},  \qquad
H = s_{\alpha} \phi^{0,r} + c_{\alpha} H_1^{0\prime},
\label{mh-mH}
\end{equation}
where we define $c_{\alpha} = \cos\alpha$, $s_{\alpha} = \sin\alpha$.
The mixing is controlled by the $2\times 2$ mass-squared matrix
\begin{equation}
\mathcal{M}^2 = \left( \begin{array}{cc}
\mathcal{M}_{11}^2 & \mathcal{M}_{12}^2 \\
\mathcal{M}_{12}^2 & \mathcal{M}_{22}^2 \end{array} \right),
\end{equation}
where
\begin{eqnarray}
\mathcal{M}_{11}^2 &=& 8 \lambda_1 v_{\phi}^2, \nonumber \\
\mathcal{M}_{12}^2 &=& \frac{\sqrt{3}}{2} v_{\phi} 
\left[ - M_1 + 4 \left(2 \lambda_2 - \lambda_5 \right) v_{\chi} \right], \nonumber \\
\mathcal{M}_{22}^2 &=& \frac{M_1 v_{\phi}^2}{4 v_{\chi}} - 6 M_2 v_{\chi} 
+ 8 \left( \lambda_3 + 3 \lambda_4 \right) v_{\chi}^2.
\end{eqnarray}
The mixing angle is fixed by 
\begin{eqnarray}
\sin 2 \alpha &=&  \frac{2 \mathcal{M}^2_{12}}{m_H^2 - m_h^2},    \nonumber  \\
\cos 2 \alpha &=&  \frac{ \mathcal{M}^2_{22} - \mathcal{M}^2_{11}  }{m_H^2 - m_h^2},    
\end{eqnarray}
with the masses given by
\begin{eqnarray}
m^2_{h,H} &=& \frac{1}{2} \left[ \mathcal{M}_{11}^2 + \mathcal{M}_{22}^2
\mp \sqrt{\left( \mathcal{M}_{11}^2 - \mathcal{M}_{22}^2 \right)^2 
	+ 4 \left( \mathcal{M}_{12}^2 \right)^2} \right].
\label{eq:hmass}
\end{eqnarray}

It is convenient to use the measured mass of the observed SM-like Higgs boson as an input parameter.  The coupling $\lambda_1$ can be eliminated in favor of this mass by inverting Eq.~(\ref{eq:hmass}):
\begin{equation}
\lambda_1 = \frac{1}{8 v_{\phi}^2} \left[ m_h^2 
+ \frac{\left( \mathcal{M}_{12}^2 \right)^2}{\mathcal{M}_{22}^2 - m_h^2} \right].
\label{eq:lambda1}
\end{equation}
Note that in deriving this expression for $\lambda_1$, the distinction between $m_h$ and $m_H$ is lost.  This means that, depending on the values of $\mu_3^2$ and the other parameters, this (unique) solution for $\lambda_1$ will correspond to either the lighter or the heavier custodial singlet having a mass equal to the observed SM-like Higgs mass.

\section{Low-$m_5$ benchmark}
\label{sec:bench}

The low-$m_5$ benchmark plane, parameterized in terms of $m_5$ (for values between 50 and 550~GeV) and $s_H$, is designed as a complementary benchmark scenario to the H5plane benchmark which covers the $m_5$ range of $[200, 2000]$~GeV. Its primary purpose is to facilitate the extension of current LHC searches to $m_5$ mass ranges below 200 GeV. In particular, since the Drell-Yan production cross section of $H^{++}H^{--}$ is independent of $s_H$, the ATLAS search for Drell-Yan production of $H^{++}H^{--}$ with decays to like-sign $W$ bosons~\cite{Aaboud:2018qcu} has the potential to exclude the entire parameter space below approximately 200 GeV if extended to masses below 200 GeV. The low-$m_5$ benchmark is specified in Table~\ref{tab:lowm5}, in a form that is compatible with the inputs of version 1.4 and higher of the GM model calculator GMCALC~\cite{Hartling:2014xma}. 

In order to establish a benchmark \emph{plane} with only two variable parameters, choices have to be made to fix the five otherwise-free parameters of the GM model.  These choices are dictated by the intended purpose of the benchmark, together with convenience of implementation.  The low-$m_5$ benchmark specifies the nine parameters of the GM model Lagrangian in terms of $m_h$, $G_F$, $m_5$, $s_H$, $\lambda_2$, $\lambda_3$, $\lambda_4$, $\lambda_5$, and $M_2$ (this set of input parameters is implemented as INPUTSET = 6 in the GMCALC code).  $m_h = 125$~GeV and $G_F$ are fixed to their measured values.

The other fixed (or dependent) parameters of the benchmark are chosen to meet the following design considerations:

\begin{itemize}
	\item The couplings of $h$ are sufficiently close to their SM values to avoid the benchmark being excluded by LHC measurements of $h$ signal strengths.  The main hazard here is that $h \to \gamma\gamma$ could be significantly modified by light $H_5^+$ and $H_5^{++}$ scalars running in the loop, in addition to (heavier) $H_3^+$ in the loop.  This possibility is conveniently avoided by the choice $\lambda_5 = -4 \lambda_2$, for which the $h H_5 H_5$ coupling strengths, given by
\begin{equation}
	g_{h55} = -8 \sqrt{3}(\lambda_3 +\lambda_4) s_\alpha v_\chi +(4 \lambda_2 + \lambda_5) c_\alpha v_\phi - 2 \sqrt{3} M_2 s_\alpha,
\end{equation}
go to zero in the limit $s_H \to 0$ (in which $s_{\alpha}$ also goes to zero).  This strong cancellation is not strictly necessary ($\lambda_2$ can be varied by up to a factor of 2 relative to this choice before Higgs signal strength measurements begin to seriously constrain the parameter space), but it is convenient and does not affect the $H_5$ production and decay phenomenology that the benchmark is designed to capture.
	\item $m_5$ is always less than $m_3$, to avoid the decays $H_5 \to H_3 V$ or $H_5 \to H_3 H_3$.  This choice is made to simplify the $H_5$ decay phenomenology in the benchmark and to emphasize the importance of the $VV$ decay modes, which are in fact dominant over the majority of the parameter space of the unconstrained GM model.  It also avoids exclusion of parts of the benchmark from direct searches for light $H_3^+$ scalars.  This mass hierarchy is achieved through the choices of the signs and magnitudes of $M_2$, $\lambda_3$, and especially $\lambda_5$.
	\item A sufficiently large portion of the $m_5$--$s_H$ plane is allowed by theoretical and indirect experimental constraints\footnote{Details of the theoretical and experimental constraints are given in the next section.} within the benchmark for $m_5$ values below 200~GeV.  This is the most challenging consideration to satisfy and dictates the specific choices for the values of $M_2$, $\lambda_3$, $\lambda_4$, and $\lambda_5$.  These choices were largely obtained by trial and error informed by knowledge of the theoretical constraints on the scalar potential.  
\end{itemize}

The remaining underlying Lagrangian parameters are given by: 
\begin{eqnarray}
 	M_1 &=& m_5^2 \frac{\sqrt{2} s_H}{c_H^2 v} - \frac{3}{\sqrt{2}} \lambda_5 v s_H - \frac{\sqrt{2}}{c_H^2} \lambda_3 v s_H^3 - \frac{6 s_H}{c_H^2} M_2, \\
      \mu_3^2 &=& \frac{2}{3} m_5^2 + M_1 \frac{v_\phi^2}{12 v_\chi} - 2  \lambda_2 v_\phi^2 - \frac{28}{3} \lambda_3 v_\chi^2 - 12 \lambda_4 v_\chi^2 - 2 M_2 v_\chi, \\
      \lambda_1 &=& \frac{1}{8 v_\phi^2} \left(m_{h}^2 
      + \frac{\left[\frac{\sqrt{3}}{2} v_\phi (-M_1 +4 (2 \lambda_2-\lambda_5)v_\chi)\right]^2}{ M_1 \frac{v_\phi^2}{4 v_\chi} -6 M_2 v_\chi + 8 (\lambda_3 +3 \lambda_4) v_\chi^2 - m_h^2}\right), \\
      \mu_2^2 &=& -4 \lambda_1 v_\phi^2 - 3 (2 \lambda_2 - \lambda_5) v_\chi^2 + \frac{3}{2} M_1 v_\chi.
\end{eqnarray}
  	
\begin{table}
\begin{tabular}{lll}
\hline\hline
Fixed inputs & Variable parameters & Dependent parameters \\
\hline
$G_F = 1.1663787 \times 10^{-5}$~GeV$^{-2}$ & $m_5 \in (50,550)$~GeV & $\lambda_2 = 0.08 (m_5/100~{\rm GeV})$ \\
$m_h = 125$~GeV & $s_H \in (0,1)$ & $\lambda_5 = -0.32 (m_5/100~{\rm GeV}) = -4 \lambda_2$ \\
$\lambda_3 = -1.5$ & & \\
$\lambda_4 = 1.5 = -\lambda_3$ & & \\
$M_2 = 10$~GeV & & \\
\hline\hline
\end{tabular}
\caption{Specification of the low-$m_5$ benchmark for the Georgi-Machacek model.  This set of input parameters can be accessed in GMCALC~\cite{Hartling:2014xma} by using INPUTSET = 6.}
\label{tab:lowm5}
\end{table}

\section{Populated range of the $m_5$--$s_H$ plane and existing constraints}
\label{sec:pop}

We now examine the populated range of the $m_5$--$s_H$ plane in the low-$m_5$ benchmark and compare it to that of a general scan of the parameter space. The results are presented as scatter plots in the $m_5$--$s_H$ plane. For all the parameter scans shown here, we impose the theoretical constraints as implemented in GMCALC~\cite{Hartling:2014xma} -- i.e., perturbative unitarity of quartic couplings~\cite{Aoki:2007ah}, a scalar potential that is bounded from below~\cite{Hartling:2014zca}, and the absence of deeper alternative minima~\cite{Hartling:2014zca}. We also impose the indirect constraints, also implemented in GMCALC, from the $S$ parameter and from $B$ physics, the most stringent in the low-$m_5$ region being $B_s^0 \rightarrow \mu^+ \mu^-$ (the constraint from $b \to s \gamma$ is very similar)~\cite{Hartling:2014aga}. We finally examine the direct experimental constraints on the low-$m_5$ region, applying direct search constraints, theory recasts and constraints from HiggsBounds 5.3.2 and HiggsSignals 2.2.3~\cite{Bechtle:2013wla}, all of which have been implemented in or interfaced to GMCALC~1.5.0~\cite{Ismail:2020zoz}.  A similar study of experimental constraints in a parameter scan of the GM model including roughly the same low mass region we are considering was previously carried out in Ref.~\cite{Chiang:2015amq}.

\begin{figure}
	\resizebox{0.7\textwidth}{!}{\includegraphics{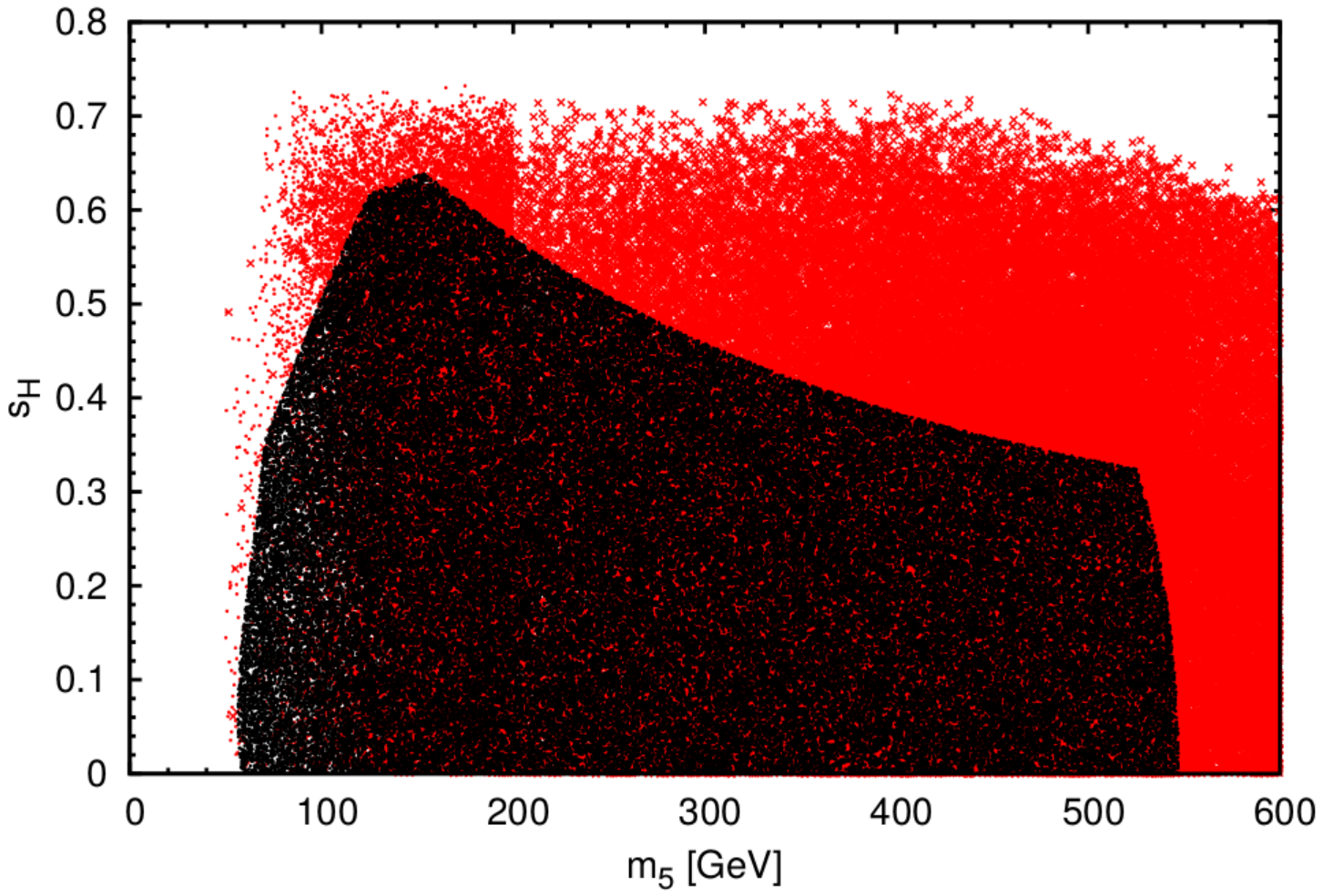}}%
	\caption{The region of $m_5$ versus $s_H$ that is populated by the low-$m_5$ benchmark (black) compared to a general scan over the full parameter space (red), imposing the same theoretical and indirect experimental constraints ($S$ parameter and $B_s \to \mu\mu$).  Because our general scan is less efficient for small $m_5$ values, we include a dedicated general scan restricted to $m_5 < 200$~GeV overlaid with a scan for $m_5 < 600$~GeV in order to generate a sufficient number of points below 200~GeV.}
	\label{fig:benchPop}
\end{figure}

In Fig.~\ref{fig:benchPop}, we show the region of the available parameter space that is accessible in the low-$m_5$ benchmark compared to a general scan over the whole parameter space after imposing the theoretical and indirect experimental constraints previously described. The red points correspond to a general scan over the GM parameter space while the black points correspond to a scan over the benchmark. As can be seen, the low-$m_5$ benchmark populates a substantial region of parameter space for $m_5$ below 200~GeV, only failing to cover the full portion of the high $s_H$ region. Because the benchmark is optimized to populate the low-$m_5$ region, at higher $m_5$ it does not cover the full range of $s_H$ obtainable in a general scan, and fails the theoretical constraints entirely for $m_5 \gtrsim 550$~GeV.  Nevertheless, the benchmark provides enough viable parameter space to be used for searches that span $m_5$ values from around 60~GeV to of order 500~GeV -- in particular, the full range where we expect a Run 2 search for Drell-Yan production of $H_5^{++} H_5^{--}$ to be sensitive.

\begin{figure}
	\resizebox{0.7\textwidth}{!}{\includegraphics{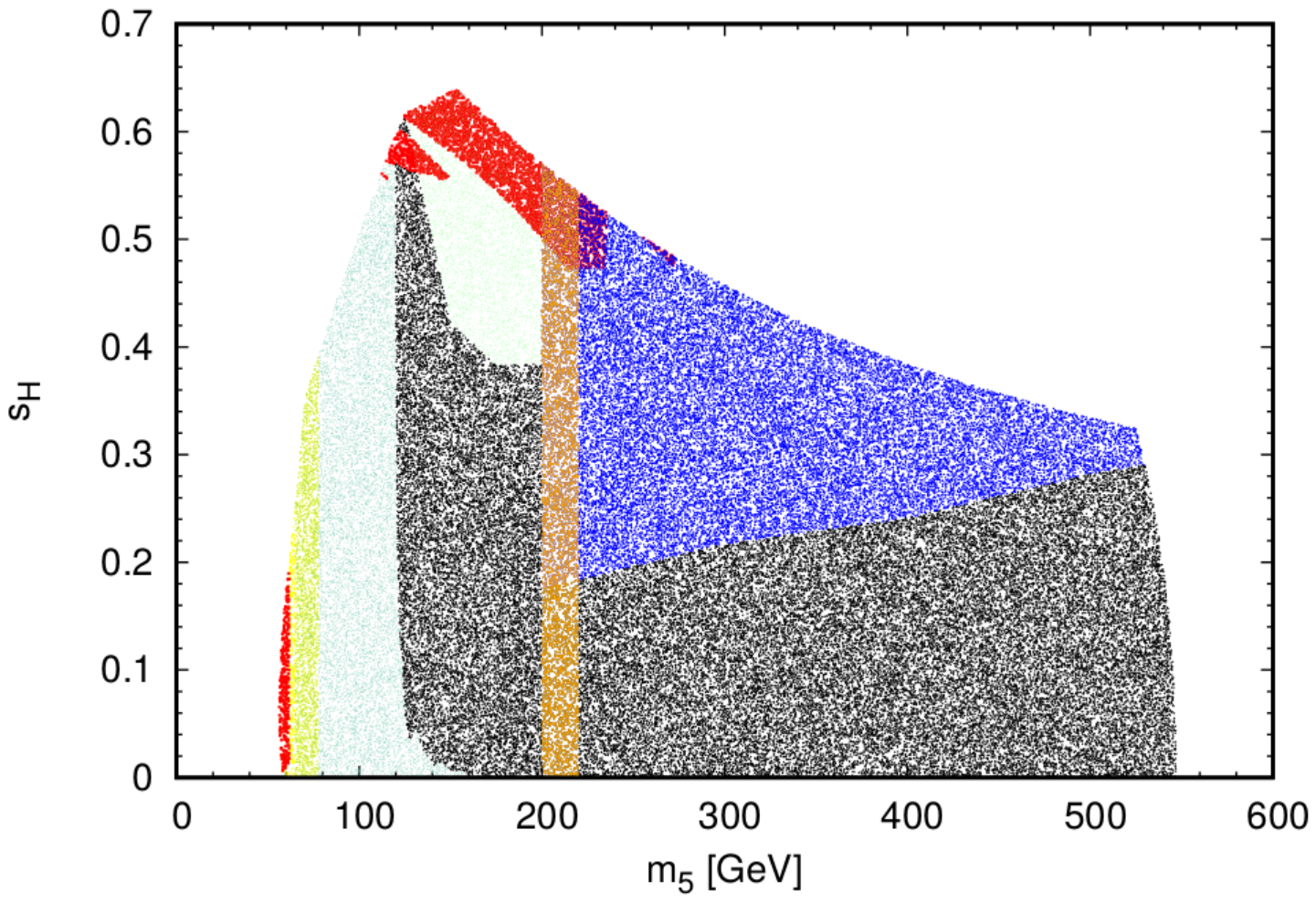}}%
	\caption{The excluded (colored) and allowed (black) points in the scan of the low-mass benchmark shown in Fig.~\ref{fig:benchPop}. The black points are unconstrained, the dark blue points are excluded by the CMS Run 2 direct search for likesign dileptons~\cite{Sirunyan:2017ret}, the orange points are excluded by the ATLAS Run 2 direct search for $H_5^{\pm\pm}$~\cite{Aaboud:2018qcu}, the light blue points are excluded by the recast of the ATLAS Run 1 diphoton resonance search~\cite{Aad:2014ioa}, the yellow points are excluded by the recast of the ATLAS Run 1 constraint on anomalous like-sign dimuon production~\cite{ATLAS:2014kca}, the teal points are excluded by the recast of the ATLAS Run 1 VBF $\to W^+W^+ \to$ like-sign dileptons cross section measurement~\cite{Aad:2014zda}, and the red points are excluded by HiggsBounds 5.3.2.}
	\label{fig:benchCon}
\end{figure}

In Fig.~\ref{fig:benchCon} we show the benchmark scan and highlight the regions excluded by the various direct experimental constraints as a scatter plot in the $m_5$--$s_H$ plane.  The black points are allowed. The exclusions come from the following sources: a CMS Run 2 search for VBF $H^{\pm\pm} \rightarrow W^{\pm} W^{\pm} \rightarrow$ like-sign dileptons~\cite{Sirunyan:2017ret}, which excludes $s_H$ values above 0.2--0.3 for $m_5 > 200$~GeV (dark blue); an ATLAS Run 2 search for Drell-Yan production of $H^{++}H^{--}$ with $H^{\pm\pm} \rightarrow W^{\pm} W^{\pm}$~\cite{Aaboud:2018qcu} which excludes  200~GeV $< m_5 < 220$~GeV (orange); an ATLAS Run 1 measurement of the VBF $\rightarrow W^+W^+ \rightarrow$ like-sign dileptons cross section~\cite{Aad:2014zda} recast by theorists in Ref.~\cite{Chiang:2014bia}, which excludes $s_H$ values above 0.4--0.6 for $m_5$ between about 120 and 200~GeV (teal); an ATLAS Run 1 search for anomalous like-sign dimuon production~\cite{ATLAS:2014kca} recast by theorists in Refs.~\cite{Kanemura:2014ipa} and~\cite{Logan:2015xpa} to constrain Drell-Yan production of $H_5^{\pm\pm}$, which excludes $m_5$ values below 76~GeV independent of the value of $s_H$ (yellow); an ATLAS Run 1 diphoton resonance search~\cite{Aad:2014ioa} applied to Drell-Yan production of $H_5^0 H_5^{\pm}$ followed by $H_5^0 \to \gamma\gamma$ which excludes points mainly for $m_5$ between 80 and 120~GeV (light blue); and constraints coming from HiggsBounds 5.3.2~\cite{Bechtle:2013wla} (red), which exclude small regions around $s_H \sim 0.45$--0.65 and $m_5 \sim 110$--230~GeV from searches for $H \to ZZ$ and a sliver with $m_5 \simeq 55$--62~GeV and $s_H \lesssim 0.2$ from searches for $h \to H_5^0 H_5^0 \to 4\gamma$. 

The CMS direct search for VBF $H_5^{\pm\pm} \to W^\pm W^\pm$  is the dominant constraint in the region above $m_5 =$ 200 GeV. The dominant constraints below $m_5 =$ 200 GeV come from processes involving $H_5^0 \to \gamma \gamma$. The dominant exclusion is from Drell-Yan production of $H_5^0 H_5^{\pm}$ with the $H_5^0$ decaying to diphotons. Because the Drell-Yan cross section does not depend on $s_H$, the excluded region is almost independent of $s_H$ (the tail extending to higher $m_5$ values at low $s_H$ is due to the increase in BR($H_5^0 \to \gamma\gamma$) as the competing decays to $WW$ and $ZZ$ are suppressed). This channel excludes $m_5$ between 65 and 120~GeV in the low-$m_5$ benchmark for any value of $s_H$. Although there is a Run 2 version~\cite{Aaboud:2017qph} of the Run 1 search for anomalous likesign dimuon production  which one would expect to compete with the diphoton exclusion, it can not be recast for our purposes because, unlike the Run 1 search, it does not consider invariant masses of the same-sign dileptons below 200~GeV. HiggsBounds exclusions from $h \to H_5^0 H_5^0 \to 4\gamma$ also exclude $m_5$ between 55--62~GeV for $s_H \lesssim 0.2$. The remaining parameter space below 65~GeV is excluded by the Drell-Yan production of $H_5^{++}H_5^{--}$ with decays to like-sign dimuons. The limits at higher $s_H$ below $m_5=$ 200~GeV are almost entirely due to VBF production of $H_5^{\pm\pm} \to W^{\pm}W^{\pm} \to$ like-sign dileptons. The HiggsBounds exclusions from production of $H$ decaying to $ZZ$ also exclude a portion of the high $s_H$, $m_5 \approx$ 200~GeV region. 

We also apply the constraints from the LHC measurements of the 125~GeV Higgs boson signal strength measurements using HiggsSignals 2.2.3~\cite{Bechtle:2013wla}.  HiggsSignals performs a p-value test given a specified number of free model parameters.  We treat each point as its own model, and hence compute the p-value with zero free parameters.  (Varying the number of free parameters in the HigssSignals analysis between 0 and 2 did not have a significant effect on the p-values.)  HiggsSignals does not exclude any of the points of the low-$m_5$ benchmark; the returned p-values vary between 0.54 and 0.75, with a p-value of 0.7 over most of the parameter space.  This is due in part to our choice of $\lambda_5 = -4 \lambda_2$, which suppresses the tree-level $h H_5 H_5$ couplings that contribute to the loop-induced $h \to \gamma\gamma$ decay.

\begin{figure}
	\resizebox{0.7\textwidth}{!}{\includegraphics{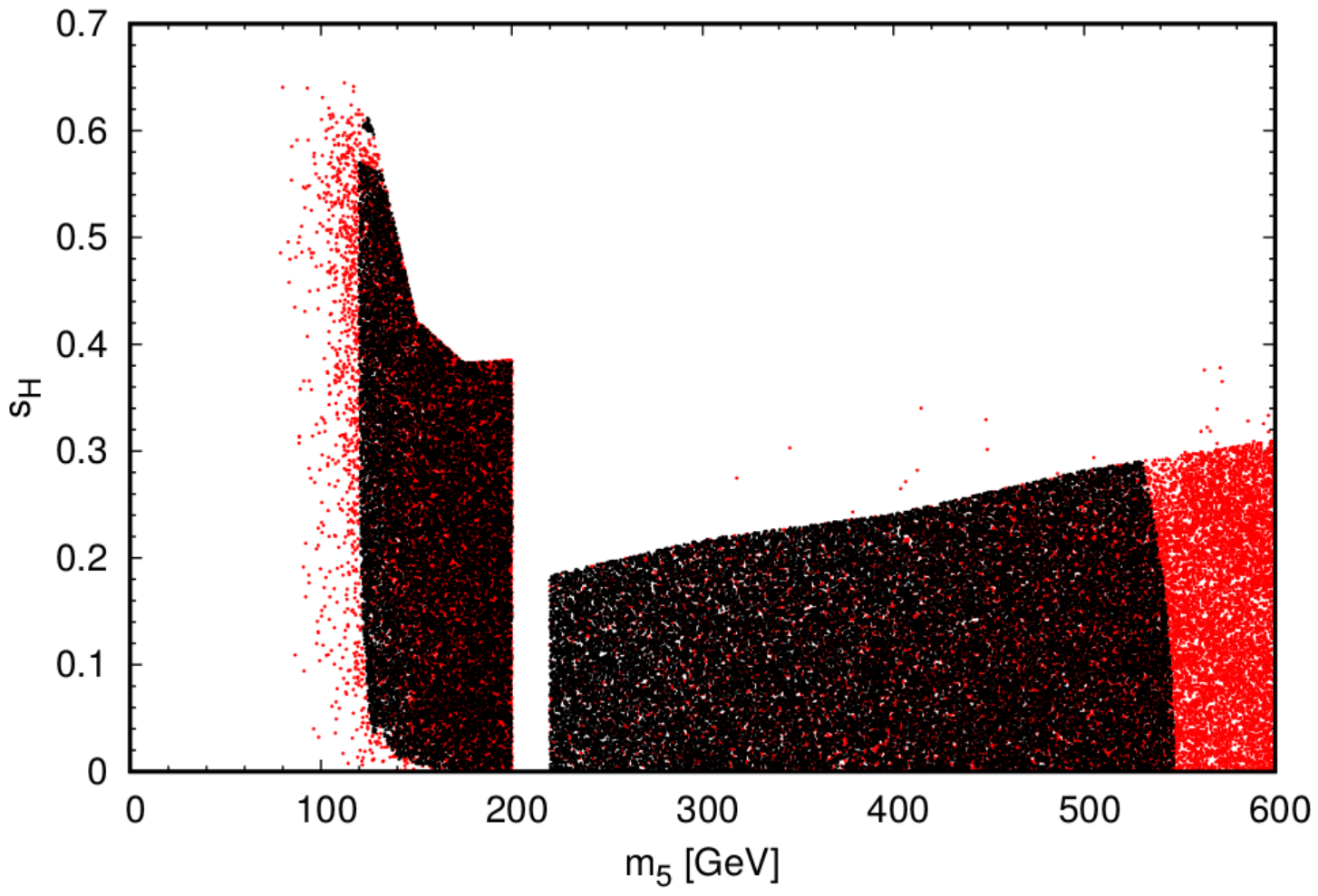}}
	\caption{The regions of the $m_5$--$s_H$ plane populated by the low-$m_5$ benchmark (black) and by a general scan (red) after all experimental constraints have been applied.}
	\label{fig:benchPopCon}
\end{figure}

In Fig.~\ref{fig:benchPopCon} we compare the coverage of the $m_5$--$s_H$ plane in the low-$m_5$ benchmark with that of the general scan after applying all the experimental constraints listed above. 
Once again, the red points correspond to a general scan, while the black points correspond to the low-$m_5$ benchmark. We can see that the low-$m_5$ benchmark can generate points covering the great majority of the allowed region. The survival of general scan points in regions excluded in the benchmark is a result of atypical suppression in the BR($H_5^{++} \to W^+ W^+$) and BR($H_5^{0} \to \gamma \gamma$) in the high and low $m_5$ region respectively. The suppression in BR($H_5^{++} \to W^+ W^+$) is caused by a non-negligible BR($H_5^{++} \to H_3^+ W^+$) while the suppression in BR($H_5^{0} \to \gamma \gamma$) is due to accidental cancellations between the gauge boson and charged scalar loop diagrams.

\section{Phenomenology}
\label{sec:pheno}

In this section we examine the properties of the low-$m_5$ benchmark. We focus on the phenomenological behaviour of the $H_5$ states and the SM-like Higgs, $h$. The results are presented as scatter plots or as contours in the $m_5$--$s_H$ plane. As above, for all the parameter scans shown here, we impose the theoretical constraints as implemented in GMCALC~\cite{Hartling:2014xma}, the indirect constraints from the $S$ parameter and from $B$ physics, and direct experimental constraints as described in the previous section.

\subsection{Couplings of $h$}

\begin{figure}
	\resizebox{0.5\textwidth}{!}{\includegraphics{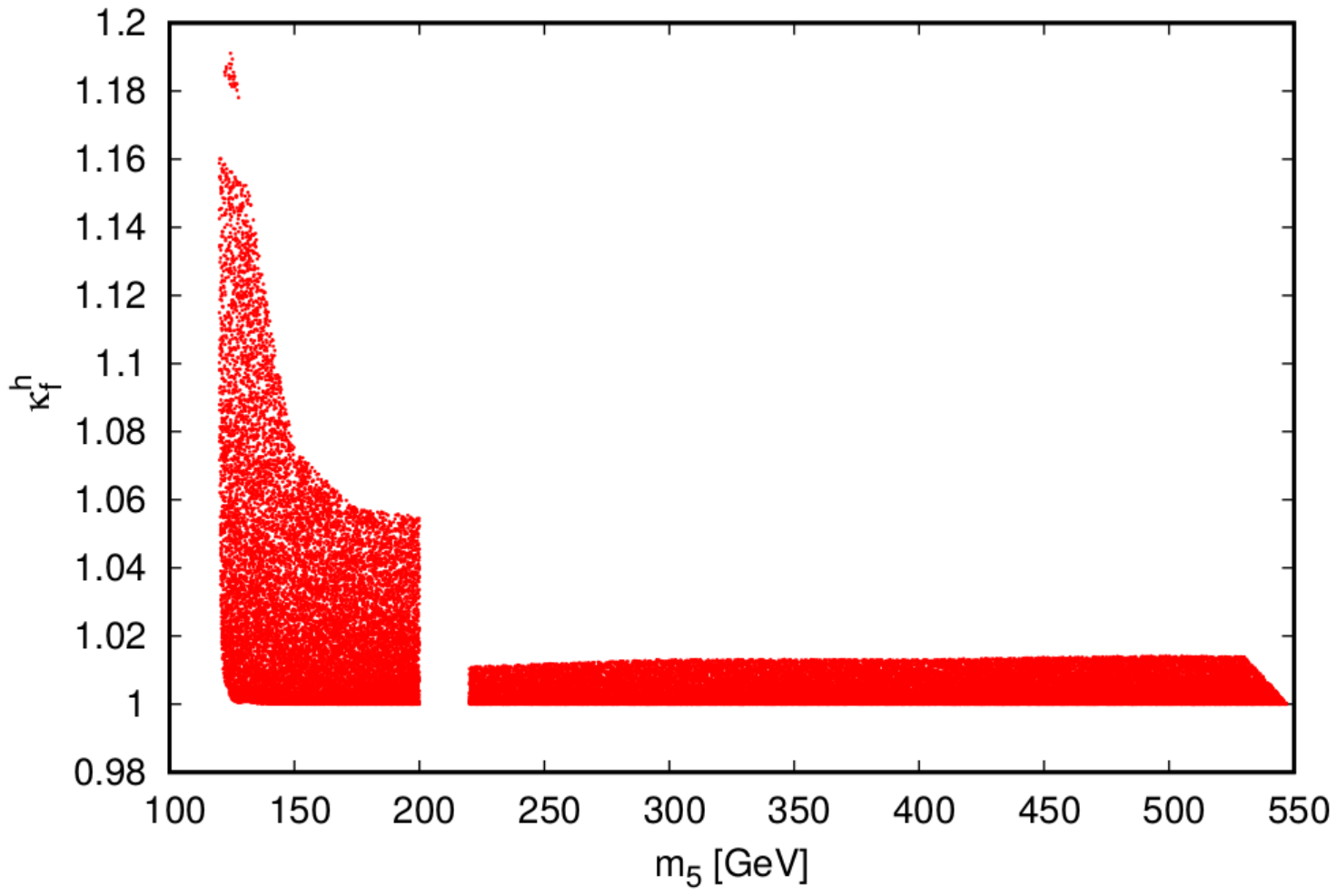}}%
	\resizebox{0.5\textwidth}{!}{\includegraphics{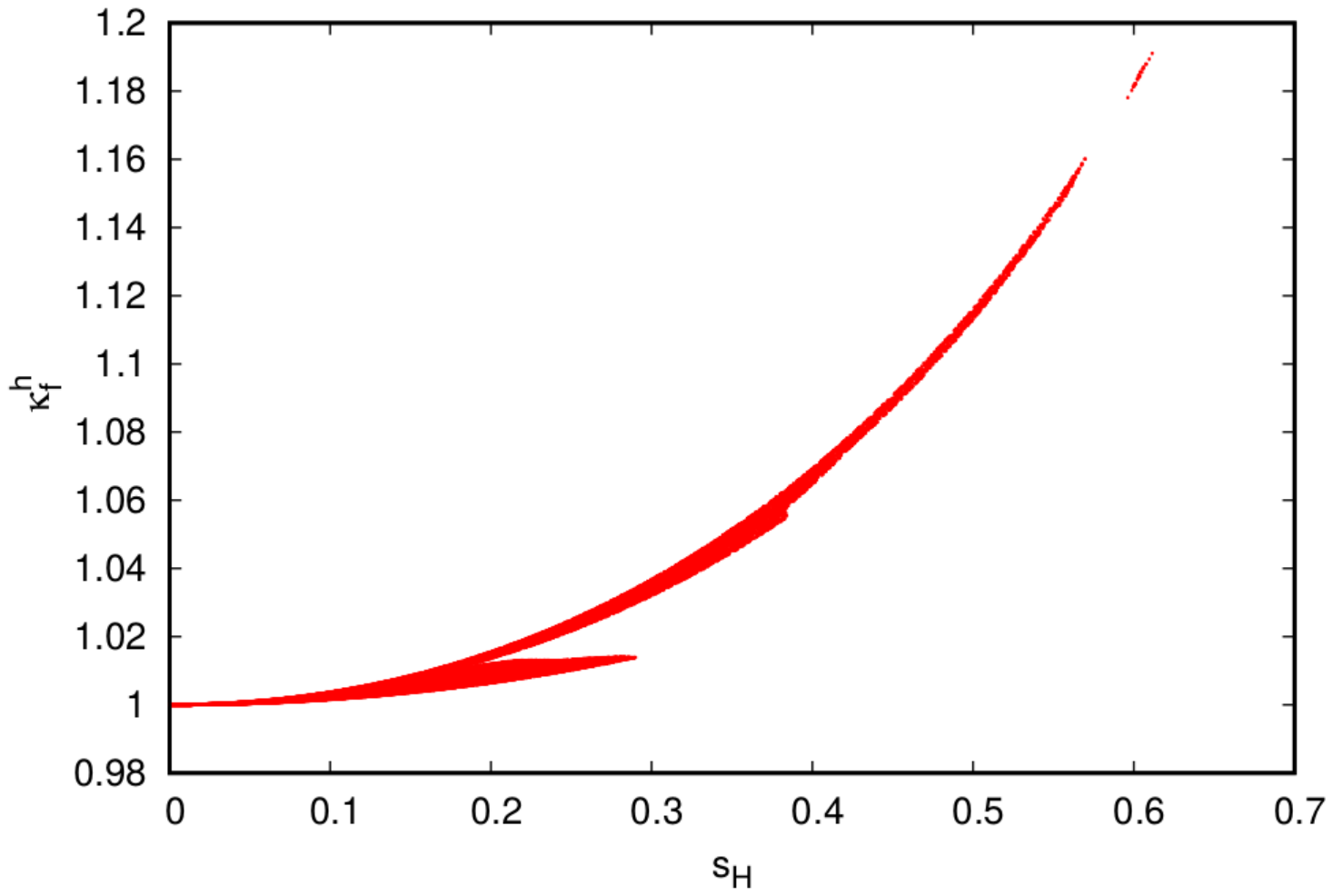}}%
	\caption{Coupling modification factor $\kappa_f^h$ for the SM-like Higgs $h$ coupling to fermion pairs in the low-$m_5$ benchmark, as a function of $m_5$ (left) and $s_H$ (right). The values range between $1$ and $1.19$.}
	\label{fig:HLKappaFL}
\end{figure}

\begin{figure}
	\resizebox{0.5\textwidth}{!}{\includegraphics{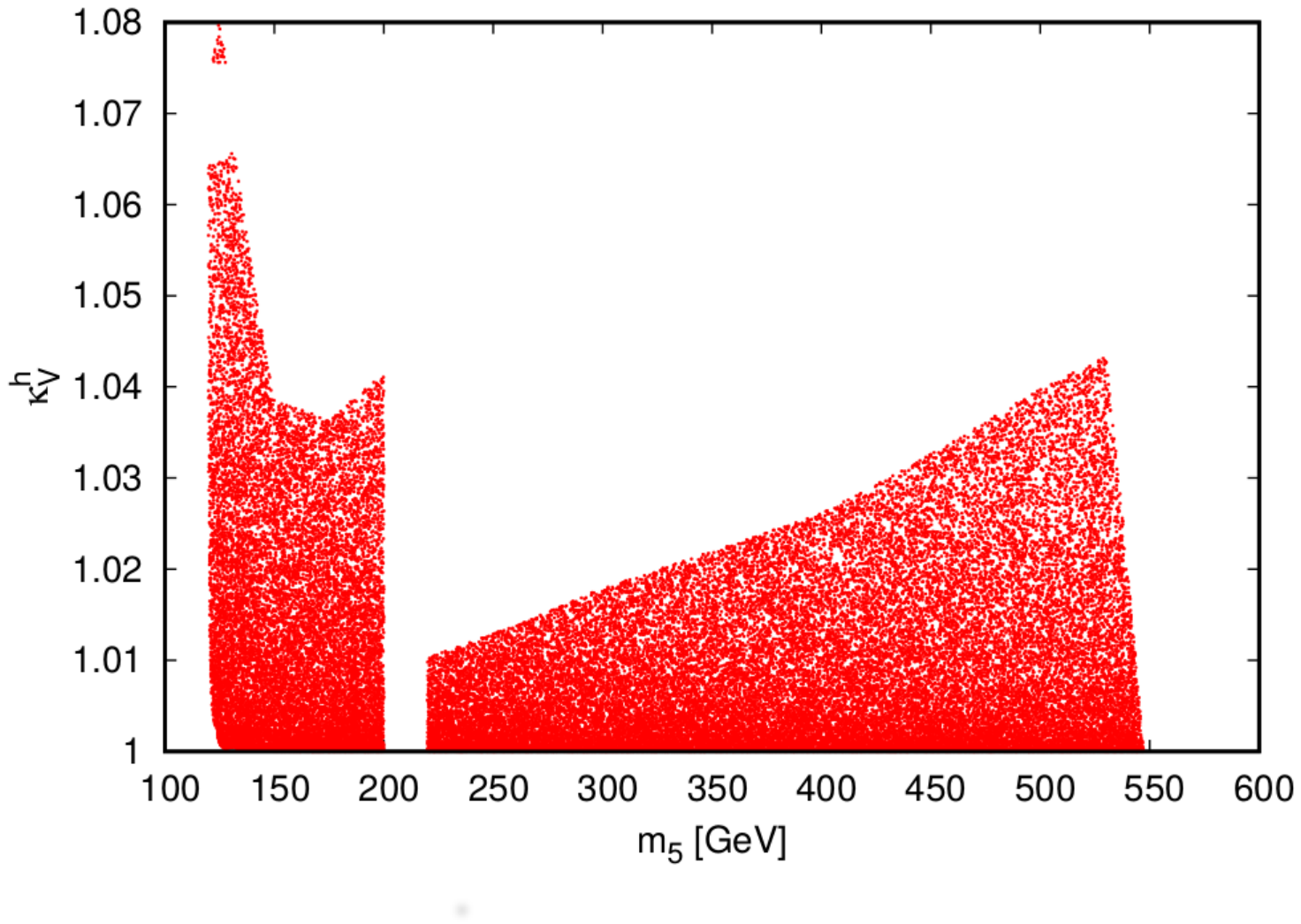}}%
	\resizebox{0.5\textwidth}{!}{\includegraphics{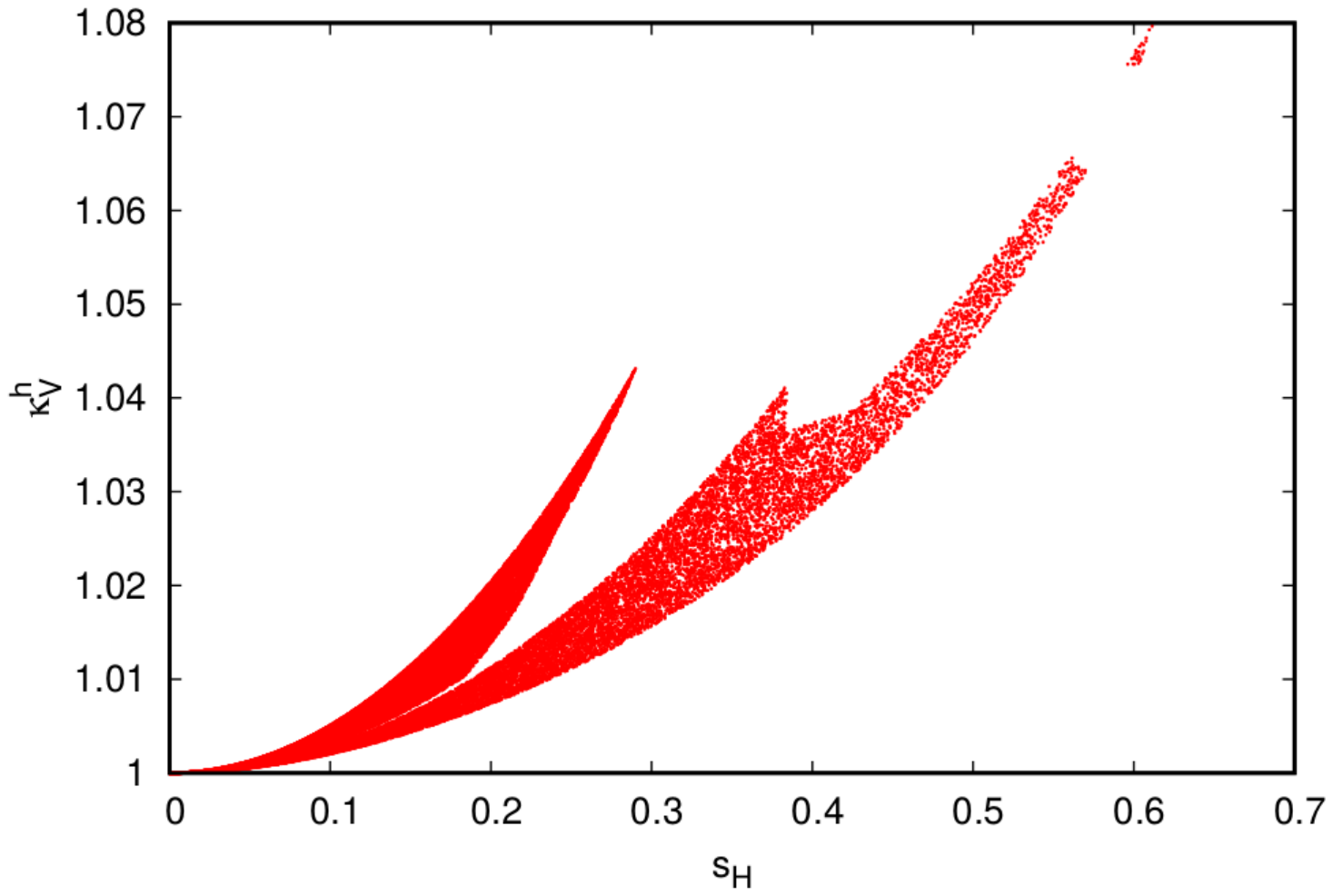}}%
	\caption{Coupling modification factor $\kappa_V^h$ for the SM-like Higgs $h$ coupling to vector boson pairs in the low-$m_5$ benchmark, as a function of $m_5$ (left) and $s_H$ (right). The values range between $1$ and $1.08$.}
	\label{fig:HLKappaVL}
\end{figure}

The tree-level couplings of the 125~GeV Higgs boson $h$ to fermion pairs and vector boson pairs are expressed in terms of model parameters by
\begin{equation}
	\kappa_f^h = \frac{c_\alpha}{c_H},  \qquad \qquad
	\kappa_V^h = c_\alpha c_H - \sqrt{\frac{8}{3}} s_\alpha s_H,
\end{equation} 
where each $\kappa$ is defined as the ratio of the corresponding $h$ coupling to that of the SM Higgs boson.  These couplings are shown in Figs.~\ref{fig:HLKappaFL} and~\ref{fig:HLKappaVL} in the low-$m_5$ benchmark, after applying all experimental constraints. Interestingly, in the benchmark both $\kappa_f^h$ and $\kappa_V^h$ are always greater than 1, so the SM-like Higgs couplings are always enhanced in the low-$m_5$ benchmark as compared to the SM (in a general scan each of these couplings can be either enhanced or suppressed relative to the SM).\footnote{It is worth noting however that this statement applies only to the tree-level couplings. Refs.~\cite{Chiang:2017vvo,Chiang:2018xpl} explored the effects of one-loop corrections to the SM Higgs couplings in the GM model and found a typical suppression of a few percent in the renormalized $\kappa_f^h$ and $\kappa_V^h$ couplings compared to their tree-level values. So we expect that at one-loop level it is possible to have $\kappa_f^h$ and $\kappa_V^h$ less than 1 in the low-$m_5$ benchmark, at least for $s_H$ values below about 0.2 for which these couplings only differ from 1 by a few percent.} The $\kappa_f^h$ enhancement can be as large as $19\%$ while the $\kappa_V^h$ enhancement can be as large as $8\%$.

In Fig.~\ref{fig:HLKappaFL} we plot $\kappa_f^h$ as a function of $m_5$ (left) and $s_H$ (right) in the low-$m_5$ benchmark. For $m_5 < 200$~GeV, a wide range of values are accessible, varying from 1 to a maximum of 1.19 at $m_5 \simeq 124$~GeV, with the maximum enhancement falling steadily to 1.06 as $m_5$ increases to 200~GeV. For $m_5 > 220$~GeV, $\kappa_f^h$ values vary from 1 to about 1.015. These two branches of the benchmark are easily distinguished in the scatter plot versus $s_H$ as the two general trends, one a rough parabola corresponding to $m_5 < 200$~GeV and the other a linear trend at lower $\kappa_f^h$ corresponding to $m_5 > 200$~GeV.  The small gap in the allowed values around $\kappa_f^h \simeq 1.17$ is a result of the small parameter region excluded by searches for $H \to ZZ$ as implemented in HiggsBounds shown in red in Fig.~\ref{fig:benchCon}.

In Fig.~\ref{fig:HLKappaVL} we plot $\kappa_V^h$ as a function of $m_5$ (left) and $s_H$ (right) in the low-$m_5$ benchmark. For $m_5 < 200$~GeV, larger enhancements of $\kappa_V^h$ are possible with values varying from 1 to a maximum of 1.08 at $m_5 \simeq 124$~GeV, with the maximum enhancement falling to below 1.04 at $m_5 \simeq 160$~GeV and then rising slightly as $m_5$ increases to 200~GeV.  For $m_5 > 220$~GeV, $\kappa_V^h$ values vary from 1 to above 1.04 with a maximum value increasing almost linearly with $m_5$.  In particular, for $m_5 > 220$~GeV, the allowed enhancements of $\kappa_V^h$ are substantially larger than those of $\kappa_f^h$.  The two branches of the benchmark are again distinguishable in the scatter plot versus $s_H$, with the branch at larger $m_5$ appearing at smaller $s_H$.  Again, the small gap in the allowed values around $\kappa_V^h \simeq 1.07$ is a result of the small parameter region excluded by searches for $H \to ZZ$ as implemented in HiggsBounds shown in red in Fig.~\ref{fig:benchCon}.  Note that $\kappa_V^h > 1$ at tree level is a distinctive feature of the GM model that cannot occur in extended Higgs sectors containing only scalars in doublets or singlets of SU(2)$_L$.

As LHC measurements of $h$ couplings become more precise we expect that they will begin to constrain the portion of the low-$m_5$ benchmark with large $s_H$ values; crucially, however, the full range of $m_5$ available in the benchmark is consistent with very SM-like couplings of $h$.

\subsection{Production of $H_5$ at the LHC}

\begin{figure}
	\resizebox{0.9\textwidth}{!}{\includegraphics{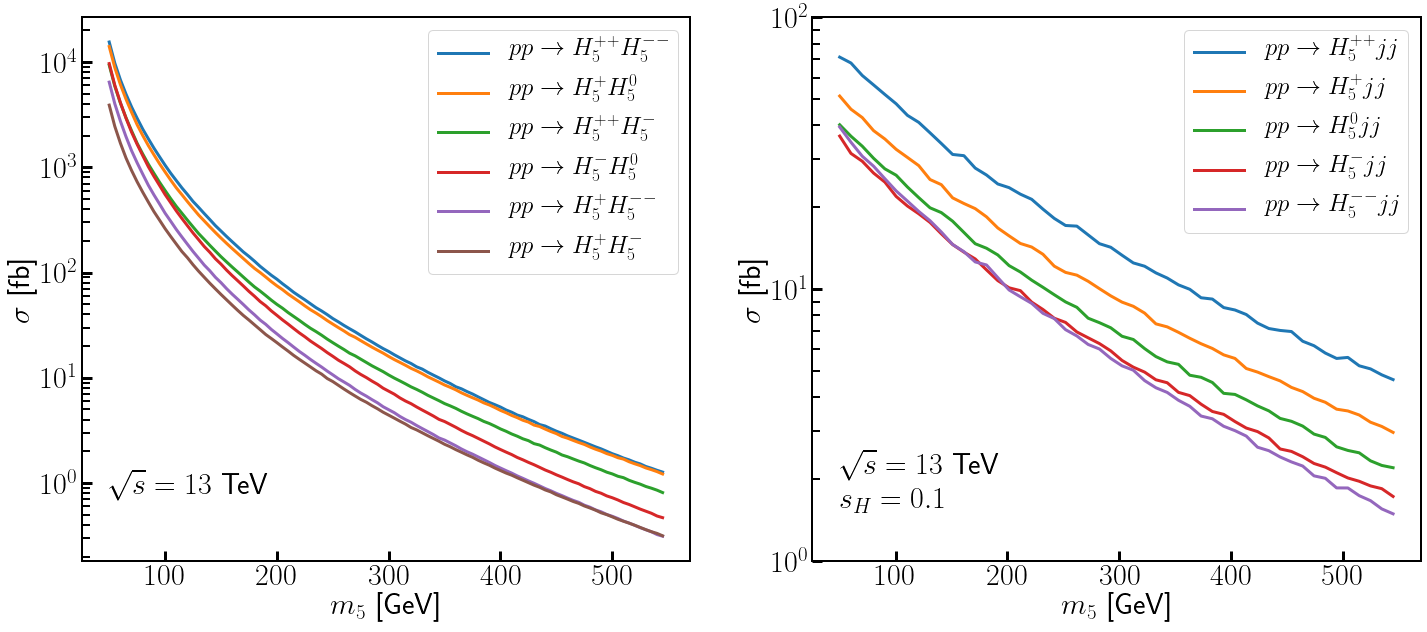}}%
	\caption{Production cross sections for Drell-Yan production of pairs of $H_5$ states (left, independent of $s_H$) and VBF production of single $H_5$ states (right, for $s_H = 0.1$) at the 13~TeV LHC, computed at NLO in QCD using MadGraph5{\_}aMC{@}NLO.}
	\label{fig:HVBFAndDY}
\end{figure}

We now discuss the major LHC production modes of the $H_5$ states in the low-$m_5$ benchmark.
In Fig.~\ref{fig:HVBFAndDY} we show the $H_5$ production cross sections via Drell-Yan production of $H_5$ pairs and via VBF. The Drell-Yan production cross section is independent of $s_H$, making it particularly valuable to constrain the GM model at low $s_H$. The VBF cross section is proportional to $s_H^2$; we fix $s_H = 0.1$ in Fig.~\ref{fig:HVBFAndDY}.  We computed these cross sections at next-to-leading order (NLO) in QCD for 13~TeV $pp$ collisions using {\tt MadGraph5{\_}aMC{@}NLO}~\cite{Alwall:2014hca} with the {\tt PDF4LHC15} NLO parton distribution functions~\cite{Butterworth:2015oua}.  VBF cross sections for production of the $H_5$ states at the LHC have previously been computed at next-to-next-to-leading order (NNLO) in QCD for $m_5 \geq 200$~GeV in Ref.~\cite{Zaro:2015ika}.  Cross sections for $H_5^{\pm\pm}$ production in VBF, associated production with a $W$ boson, and Drell-Yan production in pairs were also computed at leading order for 14~TeV and 100~TeV proton-proton collisions in Ref.~\cite{Chiang:2015amq}.

The largest Drell-Yan production cross section corresponds to the $H_5^{++} H_5^{--}$ channel which falls from 545~fb at $m_5 = 120$~GeV to 
86.3~fb at $m_5$ = 200~GeV while the smallest cross section corresponds to the $H_5^{+} H_5^{-}$ channel which falls from 138~fb at $m_5$ = 120~GeV to 
21.6~fb at $m_5$ = 200~GeV. The largest VBF production cross section (calculated at $s_H = 0.1$) corresponds to the $H_5^{++} j j$ channel which falls from 40.4~fb at $m_5$ = 120~GeV to 
23.3~fb at $m_5$ = 200~GeV while the smallest cross section corresponds to the $H_5^{--} j j$ channel which falls from 18.8~fb at $m_5$ = 120~GeV to 
9.85~fb at $m_5$ = 200~GeV.  

Within the low-$m_5$ benchmark, there is a small region of parameter space in which the mass of the heavier custodial-singlet Higgs boson $H$ satisfies $m_H > 2 m_5$, kinematically allowing the decays $H \to H_5 H_5$. Production of $H$ through, e.g., gluon fusion or VBF followed by decays to $H_5$ pairs will therefore enhance the total $H_5$ pair production rate in this region of the low-$m_5$ benchmark. As is shown below, this resonant contribution is always more than an order of magnitude smaller than the cross section from Drell-Yan production. This resonant contribution is very model-dependent, and we advocate that it be omitted when setting bounds on $H_5^{++}H_5^{--}$ production via Drell-Yan in order to allow the limits to be interpreted in as model-independent a way as possible. We determine the production cross sections of $H$ using the recommended values for beyond-the-SM Higgs bosons from the LHC Higgs Cross Section Working Group~\cite{deFlorian:2016spz} together with the appropriate $\kappa^H_f$ and $\kappa^H_V$ coupling factors. The expressions for the cross sections are given by
\begin{eqnarray}
	\sigma(pp \to H \to H_5^{++} H_5^{--}) &=& \left[ (\kappa^H_f)^2 \sigma^H_{GGF} + (\kappa^H_V)^2 ( \sigma^H_{VBF} + \sigma^H_{WH} + \sigma^H_{ZH} )  \right] \times BR(H \to H_5^{++} H_5^{--}), \\
\sigma(pp \to H \to H_5^{+} H_5^{-}) &=& \left[ (\kappa^H_f)^2 \sigma^H_{GGF} + (\kappa^H_V)^2 ( \sigma^H_{VBF} + \sigma^H_{WH} + \sigma^H_{ZH} )  \right] \times BR(H \to H_5^{+} H_5^{-}), \\
\sigma(pp \to H \to H_5^{0} H_5^{0}) &=& \left[ (\kappa^H_f)^2 \sigma^H_{GGF} + (\kappa^H_V)^2 ( \sigma^H_{VBF} + \sigma^H_{WH} + \sigma^H_{ZH} )  \right] \times BR(H \to H_5^{0} H_5^{0}),
\end{eqnarray}  
where all cross sections are evaluated at $m_H$ (we use linear interpolation between values given in Ref.~\cite{deFlorian:2016spz}), $\sigma^H_{GGF}$ is the gluon-gluon fusion cross section evaluated at NNLO + next-to-next-to-leading log (NNLL) in QCD (see Ref.~\cite{deFlorian:2016spz}), and $\sigma^H_{VBF}$, $\sigma^H_{WH}$, and $\sigma^H_{ZH}$ are the VBF and $WH$ and $ZH$ associated production cross sections, respectively.  Since $BR(H \to H_5^{++} H_5^{--}) = BR(H \to H_5^{+} H_5^{-}) = 2 BR(H \to H_5^{0} H_5^{0})$, the resonant production cross-sections for $pp \to H \to H_5^{++} H_5^{--}$ and $pp \to H \to H_5^{+} H_5^{-}$ are equal and are twice as large as that of $pp \to H \to H_5^{0} H_5^{0}$.

\begin{figure}
	\resizebox{0.5\textwidth}{!}{\includegraphics{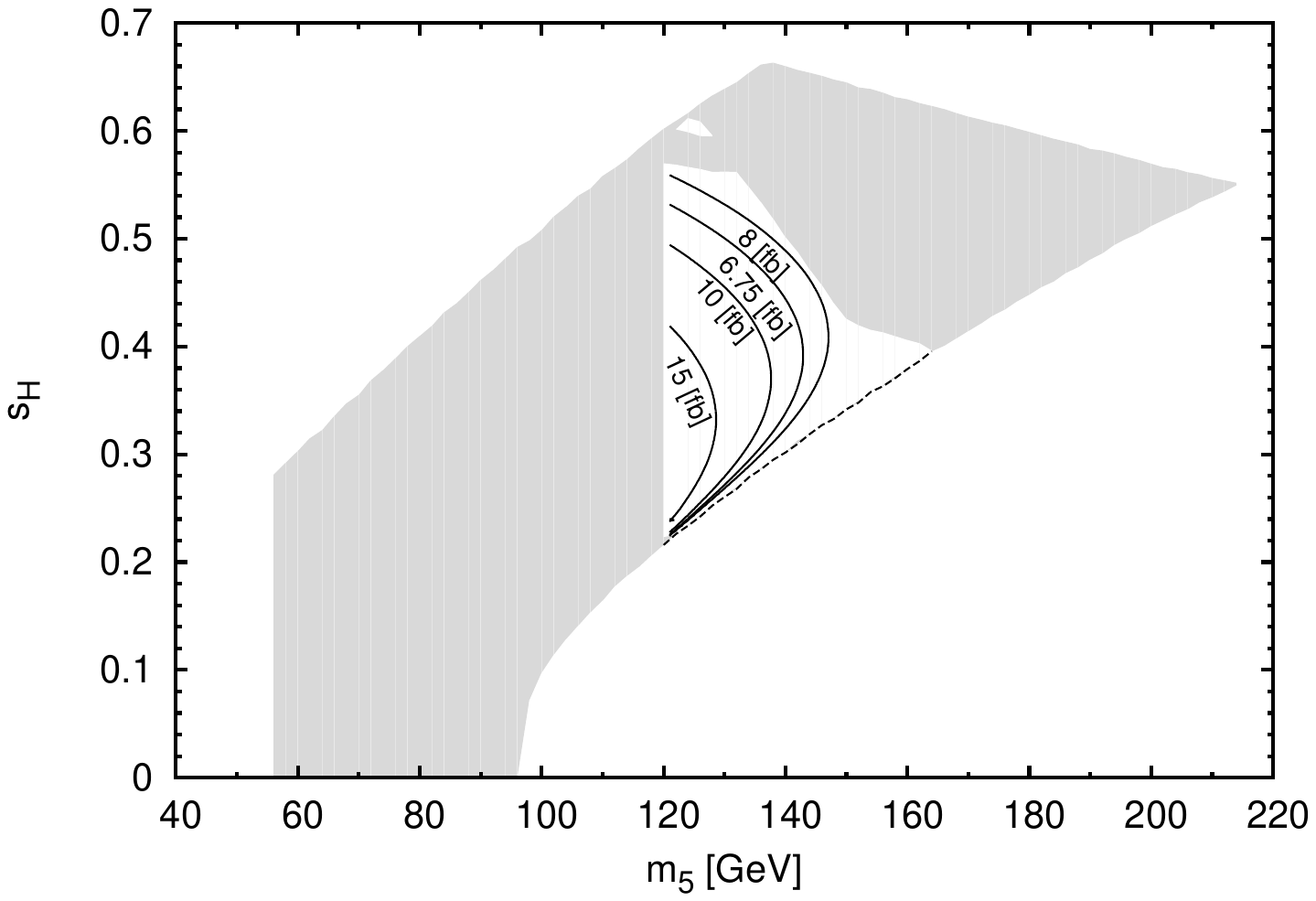}}%
	\resizebox{0.5\textwidth}{!}{\includegraphics{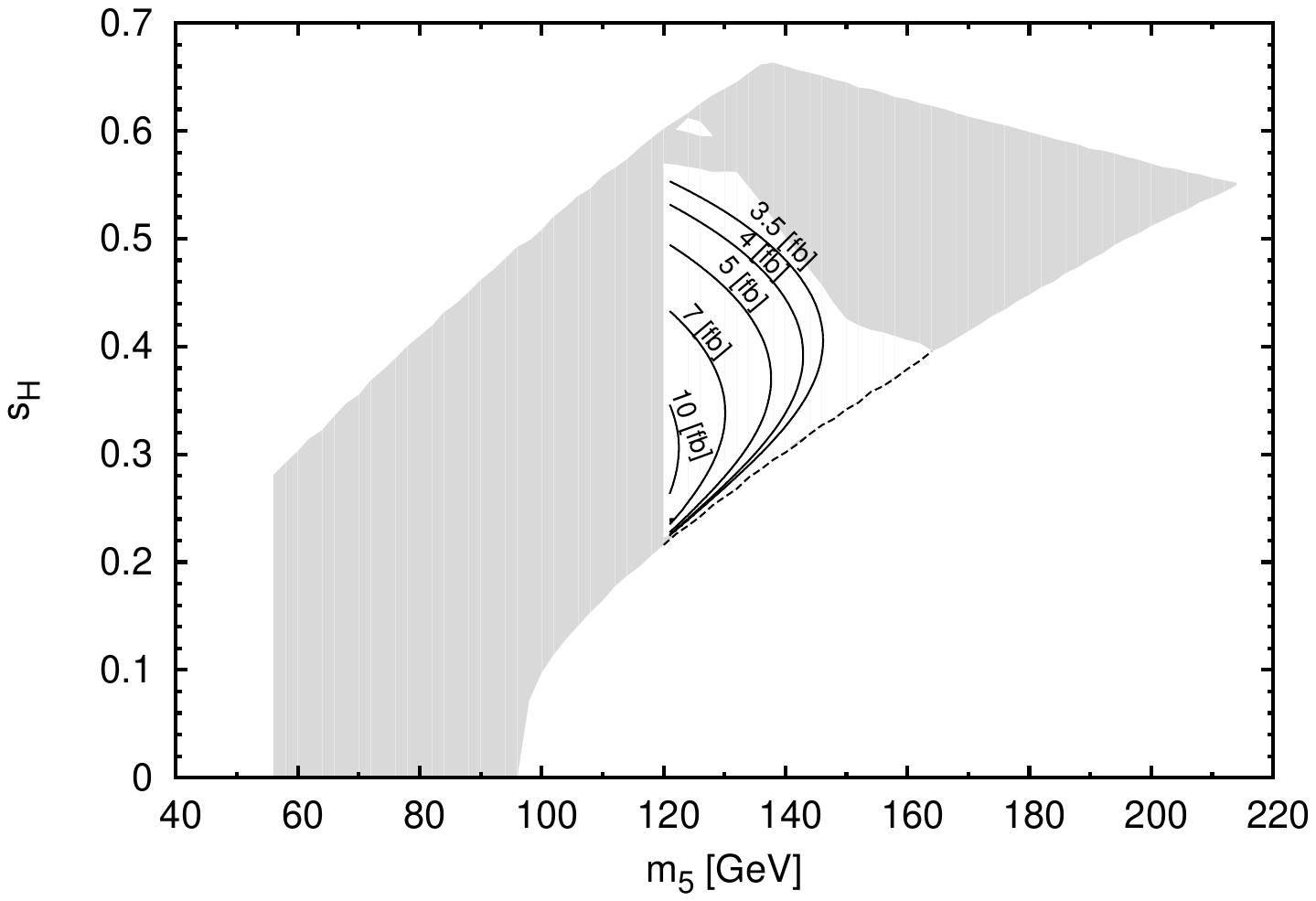}}%
	\caption{Contours of cross sections for the resonant processes $pp \to H \to H_5^{++} H_5^{--}$ (equal to $pp \to H \to H_5^{+} H_5^{-}$) (left) and $pp \to H \to H_5^{0} H_5^{0}$ (right) in the low-$m_5$ benchmark for the 13~TeV LHC. The shaded region has $m_H > 2 m_5$ but is excluded by experimental searches.  The region with contours above the dotted black line has $m_H > 2 m_5$ and is still allowed.  For $pp \to H \to H_5^{++} H_5^{--}$ and $pp \to H \to H_5^{+} H_5^{-}$, the minimum allowed cross section is 0.230~fb and the maximum allowed is 21.5~fb.  For $pp \to H \to H_5^{0} H_5^{0}$ the minimum allowed cross-section is 0.115~fb and the maximum allowed is 10.7~fb.}
	\label{fig:HXSec}
\end{figure}

In Fig.~\ref{fig:HXSec} we show the region of the $m_5$--$s_H$ plane in which $m_H > 2 m_5$ in the low-$m_5$ benchmark (gray area and white area with contours above the dotted black line).  The portion of this region shaded in gray is already excluded by existing searches.  In the surviving allowed portion of this region, which spans approximately 120~GeV $< m_5 <$ 160~GeV and $s_H$ between 0.2 and 0.6, we show contours of the cross section for $H_5$ pairs through resonant production of $H$.  
The cross sections for resonant $H_5^{++}H_5^{--}$ and $H_5^+H_5^-$ ($H_5^0H_5^0$) production have a maximum of 21.5~fb (10.7~fb) at $m_5 \approx 120$~GeV, which is more than an order of magnitude smaller than the 545~fb Drell-Yan production cross section of $H_5^{++} H_5^{--}$ at the same mass. It is also roughly an order of magnitude smaller than the individual VBF production cross sections at this mass within the relevant range of $s_H$.

\subsection{Decays of $H_5$}

The low-$m_5$ benchmark plane is designed such that $m_5 < m_3$ and $m_5 < m_H$ over the entire benchmark.  
This means that the branching ratio of $H_5^{\pm\pm} \rightarrow W^\pm W^\pm$ is 1. The decay of the $H_5^\pm$ is dominated by the $W^\pm Z$ channel for $m_5 > 170$~GeV (with a branching ratio of approximately 1) but for masses below this, the loop-induced decays to $W^\pm \gamma$ begin to compete. Similarly the decays of the $H_5^0$ are dominated by $W^+W^-$ and $ZZ$ above the appropriate thresholds but the loop-induced decays to $\gamma\gamma$ and $Z\gamma$ begin to compete at low $m_5$. In this section we show the branching ratios for the most phenomenologically interesting low-mass decays of the $H_5^+$ and $H_5^0$ to $W^+ \gamma$ and $\gamma \gamma$ respectively.  We compute these branching ratios using GMCALC, which implements doubly-offshell decays of scalars to $WW$, $ZZ$, and $WZ$ pairs below threshold as well as the loop-induced decays to $\gamma\gamma$, $Z\gamma$, and $W\gamma$~\cite{Degrande:2017naf} (decays to $Z\gamma$ and $W\gamma$ with an off-shell $Z$ or $W$ boson are not impemented).  Throughout the low-$m_5$ benchmark plane, the maximum width-to-mass ratios of the $H_5^0$, $H_5^+$, and $H_5^{++}$ are $6.4 \times 10^{-3}$, $6.5 \times 10^{-3}$ and, $6.6 \times 10^{-3}$ respectively, so the narrow-width approximation is valid throughout the benchmark. 
 
\begin{figure}
	\resizebox{0.5\textwidth}{!}{\includegraphics{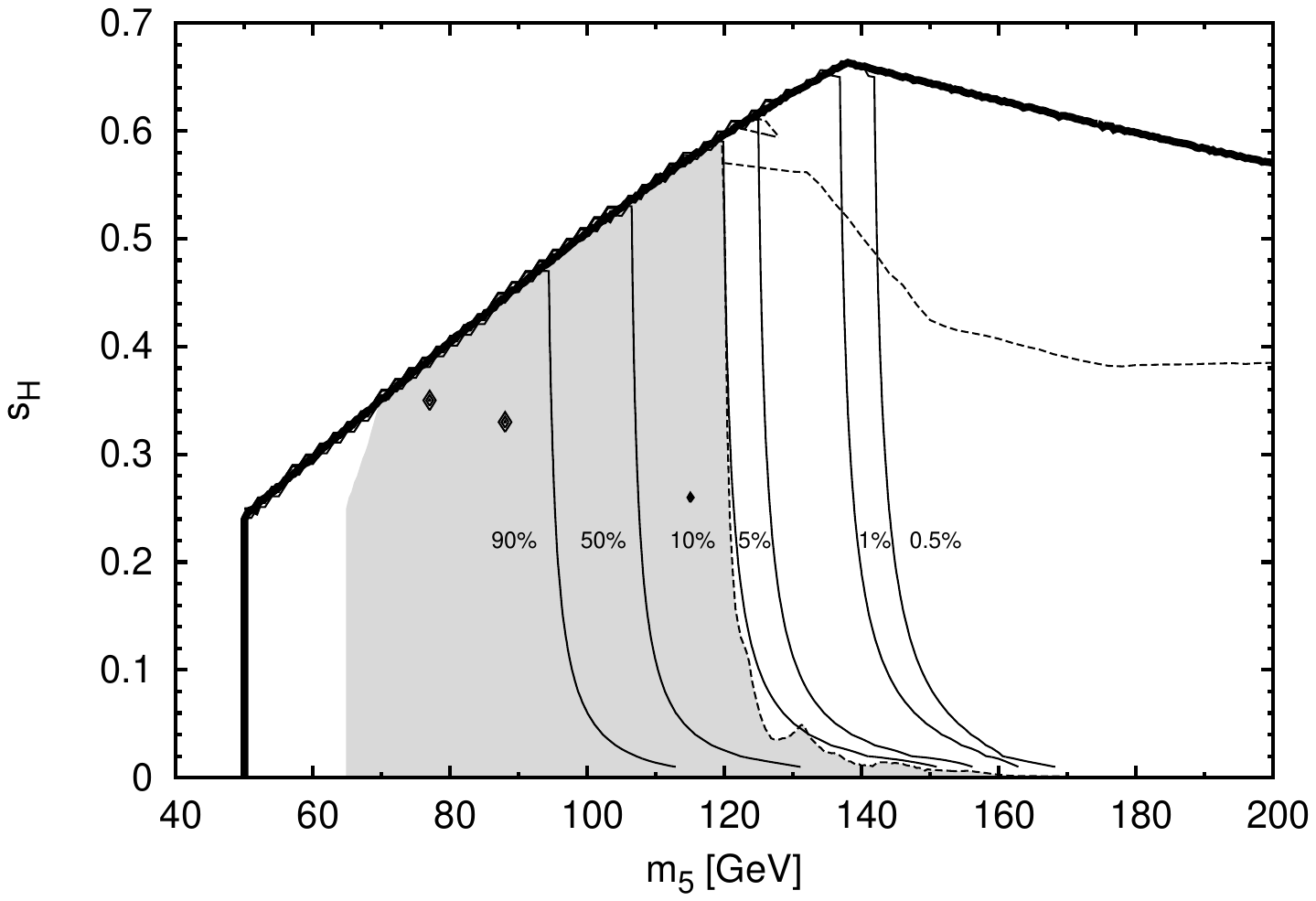}}%
	\resizebox{0.5\textwidth}{!}{\includegraphics{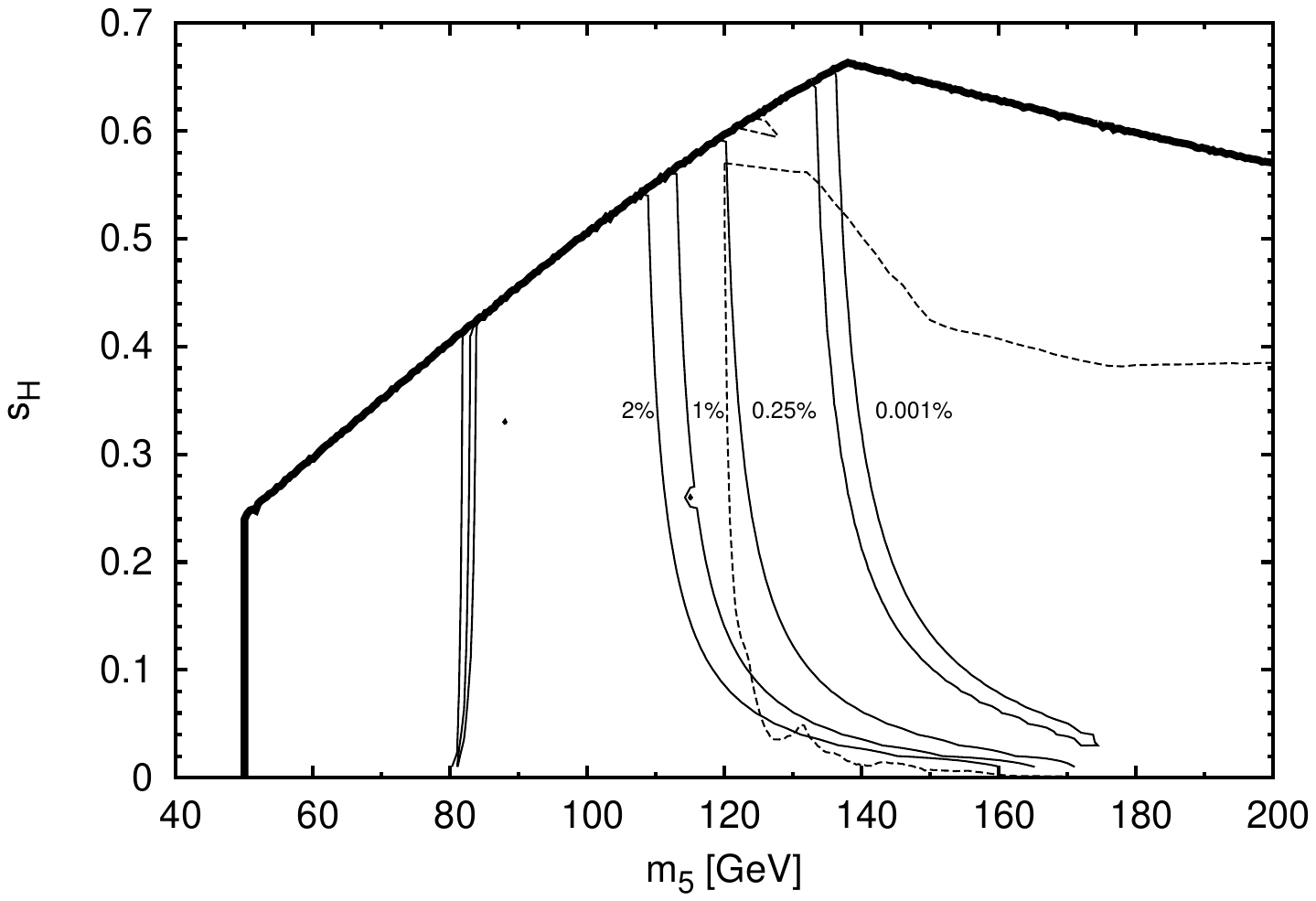}}%
	\caption{Contours of branching ratios of the loop-induced decays of $H_5^0 \to \gamma\gamma$ (left) and $H_5^+ \to W^+ \gamma$ (right) in the low-$m_5$ benchmark.  The region above and to the left of the dotted line is excluded by direct searches.  The shaded gray area in the left panel is excluded by diphoton resonance searches applied to $H_5^0 \to \gamma\gamma$. For $m_5 < 200$~GeV, the values of $BR(H_5^0 \to \gamma\gamma)$ in the low-$m_5$ benchmark range between $1.26 \times 10^{-6}$ and $1.00$, surpassing 99\% for $m_5 \lesssim 80$~GeV.  The values of $BR(H_5^+ \to W^+\gamma)$ range between zero and $0.887$ with the maximum at $m_5 \simeq 94$~GeV. }
	\label{fig:H5Decays}
\end{figure}

In Fig.~\ref{fig:H5Decays} we show the branching ratios of $H_5^0 \to \gamma \gamma$ (left) and $H_5^+ \to W^+ \gamma$ (right).  The dotted line shows the boundary of the region allowed by experimental constraints; the region below and to the right of the dotted line is allowed (as is a tiny island at $m_5 \sim 120$~GeV and $s_H \sim 0.6$). To illustrate the impact of the diphoton resonance search on the $H_5^0 \to \gamma \gamma$ decay, the region excluded by this search is shaded in gray in the right  panel of Fig.~\ref{fig:H5Decays}. Values of $BR(H_5^0 \to \gamma \gamma) \gtrsim 10\%$ are excluded since the exclusion curve closely follows the 10\% BR contour.  Values of BR($H_5^+ \to W^+ \gamma$) as large as 2\% are allowed for small values of $s_H$ below 0.05 and $m_5$ in the range 125--160~GeV.

\section{Conclusions}
\label{sec:conc}

In this paper we introduced the low-$m_5$ benchmark for the GM model, defined for $m_5 \in (50, 550)$~GeV. The benchmark highlights an unexcluded region of parameter space with light additional Higgs bosons, motivating and facilitating searches for $H_5^{++}$, $H_5^+$, and $H_5^0$ at low masses below 200~GeV. By design, $m_3 > m_5$ throughout the benchmark plane so that $BR(H_5^{++} \to W^+W^+) = 1$, while simultaneously ensuring that the modification to $h \to \gamma\gamma$  by loops of light charged scalars is suppressed. We examined the existing experimental constraints, which limit $s_H$ to be strictly below 0.63 in the benchmark, and studied its phenomenology. Our numerical work has been done with a pre-release version of the public code GMCALC~1.5.0.

We showed that the 125~GeV Higgs boson's tree-level couplings to fermion and vector boson pairs are always enhanced in the low-$m_5$ benchmark, with enhancements as large as 19\% and 8\%, respectively, possible.  Nevertheless, the 125~GeV Higgs boson satisfies the Higgs signal strength constraints throughout the entirety of the low-$m_5$ benchmark. We also showed that the benchmark-specific enhancement to the $H_5$ pair production cross sections by the process $pp \to H \to H_5 H_5$ is always at least an order of magnitude smaller than the Drell-Yan production of $H_5$ pairs. We anticipate that a dedicated experimental analysis of this Drell-Yan production at the LHC could exclude the entirety of the $m_5 < 200$~GeV region. Finally, we showed that $BR(H_5^0 \to \gamma \gamma)$ must be less than about 10\% to satisfy constraints from diphoton resonance searches, which in turn forces $BR(H_5^+ \to W \gamma)$ to be less than about 2\% in the benchmark. We also pointed out that the width-to-mass ratios of the $H_5$ states are below 1\%, small enough to justify the narrow width approximation throughout the low-$m_5$ benchmark.

\begin{acknowledgments}
We thank the members of the LHC Higgs Cross Section Working Group for encouraging us to create a benchmark valid for $m_5 < 200$~GeV.
This work was supported by the Natural Sciences and Engineering Research Council of Canada. 
A.I. was also partially supported by a Cornell Presidential Life Science Fellowship. H.E.L.\ was also supported by the grant H2020-MSCA-RISE-2014 No.\ 645722 (NonMinimalHiggs).
\end{acknowledgments}


\end{document}